\documentclass{article}

\usepackage{arxiv}

\usepackage[utf8]{inputenc} 
\usepackage[T1]{fontenc}    
\usepackage{hyperref}       
\usepackage{url}            
\usepackage{booktabs}       
\usepackage{amsfonts}       
\usepackage{nicefrac}       
\usepackage{microtype}      
\usepackage{lipsum}		
\usepackage{graphicx}
\usepackage{natbib}
\usepackage{doi}

\title{Threat-based Security Controls to Protect Industrial Control Systems}

\author{ \href{https://orcid.org/0000-0003-1064-7871}{\includegraphics[scale=0.06]{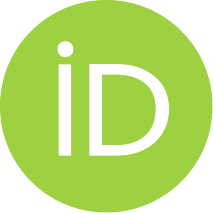}\hspace{1mm}Haritha Srinivasan}\\
	Department of Underwriting\\
	Factory Mutual Insurance Inc.\\
	Norwood, MA, 02062  \\
	\texttt{haritha.srinivasan@fmglobal.com} \\
	\And
	\href{https://orcid.org/0000-0003-2801-9080}{\includegraphics[scale=0.06]{orcid.pdf}\hspace{1mm}Maryam Karimi} \\
	Department of Innovations and Future of Industry\\
	Factory Mutual Insurance Inc.\\
	SNorwood, MA, 02062  \\
	\texttt{maryam.karimi@fmglobal.com} \\
}

\hypersetup{
pdftitle={THREAT-BASED SECURITY CONTROLS TO PROTECT INDUSTRIAL CONTROL SYSTEMS},
pdfsubject={THREAT-BASED SECURITY CONTROLS TO PROTECT INDUSTRIAL CONTROL SYSTEMS},
pdfauthor={Maryam Karimi, Haritha Srinivasan},
pdfkeywords={Adversarial Tactics, Cyber Incidents, ICS, OT, MITRE ATT\&CK, Risk Assessment, Security Reference Architecture},
}

\begin{document}
\maketitle

\begin{abstract}
This paper analyzes the reported threats to Industrial Control Systems (ICS)/Operational Technology (OT) and identifies common tactics, techniques, and procedures (TTP) used by threat actors. The paper then uses the MITRE ATT\&CK framework to map the common TTPs and provide an understanding of the security controls needed to defend against the reported ICS threats. The paper also includes a review of ICS testbeds and ideas for future research using the identified controls.
\end{abstract}


\section{Introduction}
Industrial Control Systems (ICS)/ Operational Technology (OT) interconnect, monitor, and control physical processes in a variety of industries including power generation, chemical production, manufacturing, mining, etc. ICS/OT include components such as programmable logic controllers (PLC), distributed control systems (DCS), supervisory control and data acquisition (SCADA) and human-machine interface (HMI) to control physical equipment. These systems were originally designed to run in isolated networks, with proprietary protocols. Hence, security was not part of the system design. But the introduction of commercial off-the-shelf products in ICS, and the use of information technology (IT) protocols as part of the overall ICS/OT architecture has enabled business networks to connect to these critical systems. While this facilitated better business decisions and improved efficiency, it increased vulnerability that if exploited can cause damage. 

Traditional IT security solutions cannot be applied to OT due to various operational and support differences that exist between them. OT systems include a mix of operating systems, protocols, and device types, typically for 10-20 years. The main priority of OT includes availability, safety, reliability, and then integrity, and confidentiality of digital data, whereas IT security focuses on confidentiality, integrity, and availability of digital data in that specific order. For example, patching, which is a common practice in IT environments, is not routinely done in OT environments, since patching can interrupt the availability of ICS/OT.

Another emerging trend in OT environments is the industrial internet of things (IIoT). NIST~\citep{1-industrial_iot} defines IIoT as “The sensors, instruments, machines, and other devices that are networked together and use Internet connectivity to enhance industrial and manufacturing business processes and applications.” Introduction of IIoT further increases the need for ICS security. ICS cyberattacks can lead to equipment damage, loss of process control visibility, service interruptions and even loss of life. 
To better understand and protect the OT environments and understand the need for threat-based security controls, this paper presents the security challenges of ICS/OT systems and analyzes the reported ICS threats to highlight the increasing threat landscape. The paper then identifies common tactics, techniques, and procedures (TTP) used by the threat actors with guidance from the MITRE ATT\&CK (Adversarial Tactics, Techniques, and Common Knowledge) framework~\citep{3-mclaughlin_cybersecurity} and develops the most essential security controls to defend against the reported threats. This paper also provides a review on efforts from industry and academia related to ICS testbeds and provides ideas for future research based on the MITRE framework.  An introduction to the MITRE ICS framework is also included.

\subsection{MITRE ICS ATT\&CK Framework}
The ICS ATT\&CK framework has gained much traction within the community in recent years. It originated from MITRE internal research, and, after a thorough peer review process, it was released to the public in 2020~\citep{2-mitre_ics_techniques}. ICS ATT\&CK was created to collect knowledge about malicious actors targeting the ICS environment. It was built using the already established MITRE ATT\&CK for enterprise framework, which has been widely adopted by cybersecurity teams around the world. The ICS framework, however, focusses on the ICS adversary characteristics observed in levels 0-2 of the Purdue model discussed in Section~\ref{sec2: related works}. 
As more ICS incidents started to happen, patterns such as the adversary’s TTPs have emerged. These were mapped to the framework to provide a unique view to the community regarding the methods used by the malicious actors to develop the appropriate defense mechanisms.
Tactics and Techniques provided in ICS ATT\&CK matrix~\citep{2-mitre_ics_techniques}, represent the different ways in which an attacker is trying to access and control the ICS environment. There are many ways or techniques to achieve tactic, as shown in real-world incidents, so multiple techniques exist in each tactic category.

\subsection{Structure of the paper}
Section~\ref{sec2: related works} addresses the literature review regarding the ICS security landscape. Section~\ref{sec3: arch} includes a discussion of the evolution of safety system architectures to increase the awareness of security issues in ICS/OT systems. Section~\ref{sec4: incidenst} offers a list of ICS/OT incidents to highlight the increasing threats on ICS systems. Section~\ref{sec5: controls} identifies security controls as a function of tactics and techniques; this section utilizes the MITRE framework and ICS security standards to map the common TTPs used by threat actors in the reported ICS/OT incidents and develop security controls to defend against such threats. Section~\ref{sec6: future} includes a review of ICS testbeds to show industry’s efforts to study ICS threats, along with a discussion of future research to further enhance the security controls required to protect ICS environment. Section~\ref{sec7: conlusion} concludes the paper.

\section{ICS/OT Security controls' related works}\label{sec2: related works}
Over the years, several papers have been published regarding the ICS security landscape including detailed analyses of incidents, categorization of threats, cyberattack trends and some defense countermeasures. This section provides a chronological survey of related literature. 

In~\citep{3-mclaughlin_cybersecurity}, the authors explored the ICS landscape including the general architecture of ICS and a brief history of ICS attacks. Then the paper focusses on the importance of vulnerability assessment, the benefits of ICS testbeds for such assessment, emerging threat landscapes for ICS, and software-based mitigation measures for ICS. 

In~\citep{4-ding_survey}, the authors presented an overview of security control and attack detection based on common cyberattacks in cyber-physical systems from a control theory perspective. 

In~\citep{5-rubio_cyber_defense}, the authors studied the attack vectors of a specific type of threat in ICS and provided an analysis of a detection strategy proposed in the industry and academia. 

In~\citep{6-alladi_ics_cyberattack_trends}, the authors presented some of the well-known attacks on ICS and security solutions to mitigate such attacks.  

In~\citep{7-makrakis_ics_security}, the authors made a thorough evaluation of prominent threats and attacks against critical infrastructure and ICS and provided a categorization of such threats and vulnerabilities based on several criteria. The paper also discusses ICS protocol vulnerabilities and device vulnerabilities to provide a comprehensive view of ICS attacks.

Based on the papers surveyed, it is evident that significant effort has been expended to understand the ICS security landscape. However, we have not come across any effort to identify critical security measures by analyzing the common tactics and techniques used by malicious actors.

\section{ICS Security reference architecture and controls}\label{sec3: arch}
This section provides an overview of the ICS security reference architecture that is widely used in the field. It also includes a review of the safety system architecture, as part of the overall system architecture, to better understand the need for security controls based on identified threats (see Sections~\ref{sec4: incidenst} and Sections~\ref{sec5: controls} for details).  

\subsection{Purdue model reference architecture}
The ICS security architecture is typically represented by using the Purdue Enterprise Reference Architecture, commonly referred to as the PERA model~\citep{8-mathezer_purdue_model}. It consists of six levels.

Level 0: Physical processes – Includes field devices such as sensors, actuators, valves to control the physical process like turbines, motors, process heater, etc.

Level 1: Controllers – Includes controllers such as Programmable Logic Controllers (PLC), Distributed Control Systems (DCS), Remote Terminal Units (RTU) and safety controllers. These controllers monitor input from the Level 0 sensors and manage the output to the physical process. 

Level 2: Local Supervisory – Includes control room workstations like Human Machine Interface (HMI) and Control Historian to supervise local processes.

Level 3: Site Operation – Includes systems to manage the operations across the entire plant or site such as Plant Historian, IT services for the site, engineering workstations, and other production management servers.

ICS DMZ (Demilitarized zone) – This is key part of the security model to create a boundary or perimeter separation between IT and OT. The main role of this layer is to monitor all traffic between IT and OT, thereby eliminating direct communication between these levels. Typical services included are remote access server and Shared Historian.

Level 4: Business Networks – Includes IT services for the local plant or site such as production scheduling systems like Enterprise Resource Planning (ERP), Manufacturing Execution Systems (MES), print servers, and other business systems. 

Level 5: Enterprise Networks – Include corporate IT services like internet access points, email servers, web servers, and other enterprise systems.

\subsection{Safety system architecture}
Process control systems shown in Level 1 (PLC, RTU and DCS in Fig. 1) manage equipment, production and processes in a facility based on a preset condition. Safety systems (Level 1 controllers in Fig. 1) are used to prevent failures or unsafe conditions. Safety systems operate independently from process control systems to detect and prevent dangerous events. 

Due to the rapid development of vendor solutions that offer varying levels of integration of safety system functions with the process control network, the LOGIIC (Linking the Oil and Gas Industry to Improve Cybersecurity) consortium~\citep{9-logiic_sis_integration}, and IEC 61508 (Functional safety of electrical/ electronic/ programmable electronic safety-related systems)~\citep{10-61508_association_guidance} evaluated the security of the different safety system architectures used in the industry. They also provided recommendations to help reduce the cyber risk. Based on the articles published by LOGIIC, and IEC 61508, process control system and safety system architecture are often designated under the following categories:

1)	Air-Gapped architecture includes physically separate and independent process control network and safety system network. 

2)	Interfaced architecture retains the separate logic solver and network for the process and safety system, but typically includes a data connection between the process controller and the safety controller usually via a simple serial protocol or other vendor proprietary protocol. 

3)	Integrated architecture retains the separate controller or logic solver for the process control system and the safety system but shares a common network. Typically, the controllers are from the same vendor for easier integration. In some cases, the engineering workstation for both PCS and safety system could be shared, therefore increasing the access to the safety system and commonality of engineering tools. 

4)	Common architecture includes a shared logic solver for PCS and safety system functionality and a shared engineering workstation as well.

\subsection{Analysis of the architecture}
Air-gapped and interfaced architectures are declining in use because they either lack data exchange between the systems and/or it is difficult to create communication links between different vendor systems due to proprietary protocols and software. Hence, integrated and/or common architectures are becoming more prevalent in the industry. However, risks related to cybersecurity increase with increased use of such architectures. Even though integrated/common architectures provide cost savings because of shared platforms and easy data transfer between systems, the major risk to these architectures is that an attacker who gains access to the control system network may be able to attack the safety system and cause physical equipment damage. The risk further increases when combined with remote access, which is becoming a common practice in the industry. An attacker may use the remote access path to infiltrate the control network (e.g., HMI) and then pivot to the shared engineering workstation to make changes to the safety logic to cause physical damage to the process equipment. Such tactic was evident in the TRITON/TRISIS attack~\citep{11-dragos_trisis} in 2017 and has been a major concern. The general belief that, even if the process control system is compromised, the safety system will prevent damage has been challenged by this incident. 

Due to the prevalence of a variety of safety system architectures implemented across the industry, a careful approach to OT security is warranted, as a one-size fits all solution is insufficient. It is important to analyze the real-world incidents as shown in Section IV and adopt a threat-based approach to prevent and mitigate cyberattacks.

\section{ICS/OT Incidents}\label{sec4: incidenst}
A variety of malware attacks on ICS/OT systems, some of which are highly sophisticated, has been observed in the recent past. Most cyber-attacks on OT systems have originated through IT systems. A comprehensive study of such ICS incidents affecting critical infrastructure has been published~\citep{12-hemsley_history_ics}.  Attack events have been identified under two categories: 

1)	Targeted OT: Resulting in physical damage and/or service interruption like power loss. 

2)	Indirect OT: Due to targeted IT attacks, where OT services were disrupted, such as ransomware attacks on corporate IT environment.  When a ransomware impacts IT network, it has become a common practice for companies to disconnect OT from IT as a measure to prevent spreading of the virus. However, it impacts production since OT relies on IT technology for many manufacturing services like shipping, customer information, and inventory details. 

In the following an overview of all major ICS incidents is provided with the objective to understand targets and technique used in these attacks starting from initial access vector. 
 
\begin{itemize}  
    \item Maroochy Shire (2000, Targeted OT)~\citep{12-hemsley_history_ics}  
    \begin{itemize}  
        \item Initial access: Wireless Radio Access - Insider threat  
        \item Affected: Wastewater treatment plant SCADA system.  
        \item Result: Serious environment damage by releasing 265k gallons untreated sewage.  
    \end{itemize}  
      
    \item Slammer worm (Davis-Besse nuclear plant) (2003, Indirect OT)~\citep{13-kesler_nuclear_facilities_vulnerability}  
    \begin{itemize}  
        \item Initial access: Remote access  
        \item Affected: MS SQL server and then process control network.  
        \item Result: Inaccessible Plant's Safety Parameter Display System (SPDS) for 5 hours.  
    \end{itemize}  
  
    \item Stuxnet (Iran’s Natanz nuclear plant) (2010, Targeted OT)~\citep{12-hemsley_history_ics}  
    \begin{itemize}  
        \item Initial access: USB  
        \item Affected: Siemens PLC, first known malware to cause damage to ICS.  
        \item Result: Ruined up to 10\% of the 9,000 centrifuges in Natanz.  
    \end{itemize}  
  
    \item Shamoon (Saudi Aramco) (2012, Indirect OT)~\citep{12-hemsley_history_ics}  
    \begin{itemize}  
        \item Initial access: Insider or Phishing  
        \item Affected: Overwriting files on the hard disks of about 30,000 Windows-based workstations.  
        \item Result: IT network interruption for 2 weeks. IT systems critical to OT such as payment systems, shipment and inventory tracking systems were down.  
    \end{itemize}  
  
    \item New York Dam (2013, Targeted OT)~\citep{12-hemsley_history_ics}  
    \begin{itemize}  
        \item Initial access: Internet accessible device  
        \item Affected: SCADA system of a dam through a cellular modem.  
        \item Result: Unsuccessful attack, Attacker got remote access to information about the sluice gate's status and operation which was offline for maintenance.  
    \end{itemize}  
  
    \item German Steel Mill (2014, Targeted OT)~\citep{12-hemsley_history_ics}  
    \begin{itemize}  
        \item Initial access: Spear phishing  
        \item Affected: Disabled the ability to shut down a blast furnace.  
        \item Result: Significant physical damages.  
    \end{itemize}  
  
    \item BlackEnergy (Ukraine power grid) (2015, Targeted OT)~\citep{12-hemsley_history_ics}  
    \begin{itemize}  
        \item Initial access: Spear phishing  
        \item Result: Power outages for approximately 225,000 customers in three provinces for about six hours.  
    \end{itemize}  
  
    \item Industroyer/CrashOverride (2016, Targeted OT)~\citep{12-hemsley_history_ics}  
    \begin{itemize}  
        \item Initial access: Spear phishing  
        \item Affected: Transmission substations in Ukraine.  
        \item Result: Power outages for one hour, affecting one-fifth of the Kiev region.  
    \end{itemize}  
  
        \item “Kemuri” Water company (2016, Targeted OT)~\citep{12-hemsley_history_ics}  
    \begin{itemize}  
        \item Initial access: Internet accessible device  
        \item Affected: Water district’s valve and flow control application.  
        \item Result: Alter the amount of chemicals entering the water supply.  
    \end{itemize}  
  
    \item NotPetya (2017, Indirect OT)~\citep{12-hemsley_history_ics}  
    \begin{itemize}  
        \item Initial access: Supply chain  
        \item Affected/Result: Tens of thousands of systems in more than 65 countries, including organizations like Maersk, FedEx, Mondelez, Merck, and Reckitt Benckiser. Pharmaceutical giant, Merck’s total gross value reported was \$1.4B including sales losses and manufacturing and remediation-related expenses.  
    \end{itemize}  
  
    \item TRITON/TRISIS/HatMan (2017, Targeted OT)~\citep{11-dragos_trisis}  
    \begin{itemize}  
        \item Initial access: Corporate IT network/Engineering workstation  
        \item Affected: Gained remote access to an SIS engineering workstation and communicated with the SIS controller using a proprietary protocol.  
        \item Result: Accidental system shutdown.  
    \end{itemize}  
  
    \item VPNFilter (2018, Indirect OT)~\citep{14-us_cert_vpnfilter}  
    \begin{itemize}  
        \item Initial access: Small office/home office routers and NAS  
        \item Affected/Result: Routers and OT network-attached storage devices from various manufacturers by stealing website credentials, and monitoring Modbus and SCADA protocol.  
    \end{itemize}  
  
    \item LockerGoga (Norsk Hydro) (2019, Targeted OT)~\citep{15-austin_norsk_hydro}  
    \begin{itemize}  
        \item Initial access: Corporate IT systems  
        \item Affected: Company’s global network of over 3k servers and 1k more PCs.  
        \item Result: Forcing the company to move to manual operations, resulting in an estimated \$70 million in losses.  
    \end{itemize}  
  
    \item Colonial Pipeline (2021, Indirect OT)~\citep{16-morse_colonial_pipeline}  
    \begin{itemize}  
        \item Initial access: Corporate IT systems /VPN connection  
        \item Affected: Pipeline’s IT network (supplying half of the fuel for US East coast).  
        \item Result: Shutdown of operations for a week and affected the supply of gas.  
    \end{itemize}  
  
    \item JBS (2021, Indirect OT)~\citep{17-gatlan_jbs_foods}  
    \begin{itemize}  
        \item Initial access: Corporate IT systems  
        \item Affected: JBS’ North American and Australian IT systems.  
        \item Result: Shutdown of production at multiple sites worldwide.  
    \end{itemize}  
  
    \item Vestas (2021, Indirect OT)~\citep{18-toulas_vestas}  
    \begin{itemize}  
        \item Initial access: Corporate IT systems  
        \item Affected: Corporate IT systems.  
        \item Result: Network shutdown, forced some factories to slow down production.  
    \end{itemize}  
  
    \item Toyota/Kojima (2022, Indirect OT)~\citep{19-reuters_toyota}  
    \begin{itemize}  
        \item Initial access: Supply chain/IT systems  
        \item Affected: Halted 28 production lines across Toyota’s 14 factories because of a system malfunction at Kojima, a key supplier of parts.  
        \item Result: Halted production of 13k vehicles or 5 percent of monthly output.  
    \end{itemize}  
  
    \item Bridgestone (2022, Indirect OT)~\citep{20-nelson_bridgestone}  
    \begin{itemize}  
        \item Initial access: Supply chain/IT systems  
        \item Result: Shutdown of the network and production at Bridgestone’s factories in North and Middle America for about a week. Bridgestone is another supplier for Toyota resulting in a second impact for Toyota.  
    \end{itemize}  
  
    \item Denso (2022, Indirect OT)~\citep{21-toulas_denso}  
    \begin{itemize}  
        \item Initial access: Supply chain/IT systems  
        \item Affected: Ransomware on multibillion dollar supplier to key automotive companies like Toyota.  
        \item Result: Leaked information on purchase orders, emails, NDAs, technical drawings related to production and OT systems.  
    \end{itemize}  
  
    \item Wind Turbine Attacks (2022, Indirect OT)~\citep{22-stupp_wind_energy}  
    \begin{itemize}  
        \item Initial access: Corporate IT systems  
        \item (1) Affected: Deutsche Windtechnik, wind turbines maintenance. Result: Loss of remote control of 2000 wind turbines for a day.  
        \item (2) Affected: Ransomware on Nordex, a wind turbine maker’s IT system. Result: Shut down and disabled remote access to OT from Nordex IT.  
        \item (3) Affected: Enercon, wind turbine manufacturer, by an attack on a satellite company, ViaSat. 
        \item Result: Loss of remote control of 5800 wind turbines in central Europe, totaling 11GW.  
    \end{itemize}  
  
    \item Predatory Sparrow (2022, Targeted OT)~\citep{23-bahwe_iranian_steelmaker}  
    \begin{itemize}  
        \item Initial access: Third party software  
        \item Affected: Khouzestan Steel Company.  
        \item Result: Physical damage of an industrial machine leading to molten steel across the factory floor.  
    \end{itemize}  
  
    \item Mining Industry Attack (2022, Indirect OT)~\citep{24-seidler_aurubis, 25-clausen_copper_mountain}  
    \begin{itemize}  
        \item Initial access: Corporate IT systems  
        \item (1) Affected: IT systems in Aurubis, Europe's biggest copper smelter. 
        \item (1) Result: Forcing shutdown and disconnect from internet.  
        \item (2) Affected: Ransomware on IT systems in Canadian mining company, Copper Mountain. 
        \item (2) Result: The company with isolated operations switched to manual processes and shut down the mill to assess the effect on the control system.   
    \end{itemize}  
\end{itemize}

\section{THREAT-BASED SECURITY CONTROLS}\label{sec5: controls}
In this section, the common TTPs used by threat actors are mapped to ICS security controls to defend against the threats. A background on MITRE ATT\&CK framework and ISA/IEC 62443 standard is provided in the following.

\subsection{MITRE ATT\&CK framework development}
Prior to the ATT\&CK framework launch, it was reviewed by several private and public entities including cyber intelligence and security companies that focus on ICS~\citep{26-mitre_framework}. 
Since the release of the framework, several ICS threat intelligence companies have started mapping the threat actors’ behavior to the MITRE ICS framework. For example, Dragos has observed and reported on 20 different threat groups that are primarily targeting ICS environments around the world. Dragos mapped their findings to the ICS MITRE framework for public consumption~\citep{27-dragos_mitre_attack}]. The ICS Advisory Project~\citep{28-isa_iec_62443} released an interactive dashboard using publicly available threat/vulnerability data such as CISA ICS Advisories and MITRE ATT\&CK.  

\subsection{ISA/IEC 62443}
International Electrotechnical Commission (IEC) 62443 is an international series of standards developed to secure IACS (industrial automation and control systems). It was originally developed by ISA 99 (International Society of Automation) committee~\citep{28-isa_iec_62443}. This is one of the highly recognized security standards for ICS in the industry and details all security controls required to protect the ICS environment. It has various parts and includes nine standards, technical reports, and technical specifications.

\subsection{Methodology}
\begin{figure}
	\centering
	\includegraphics[width=\textwidth]{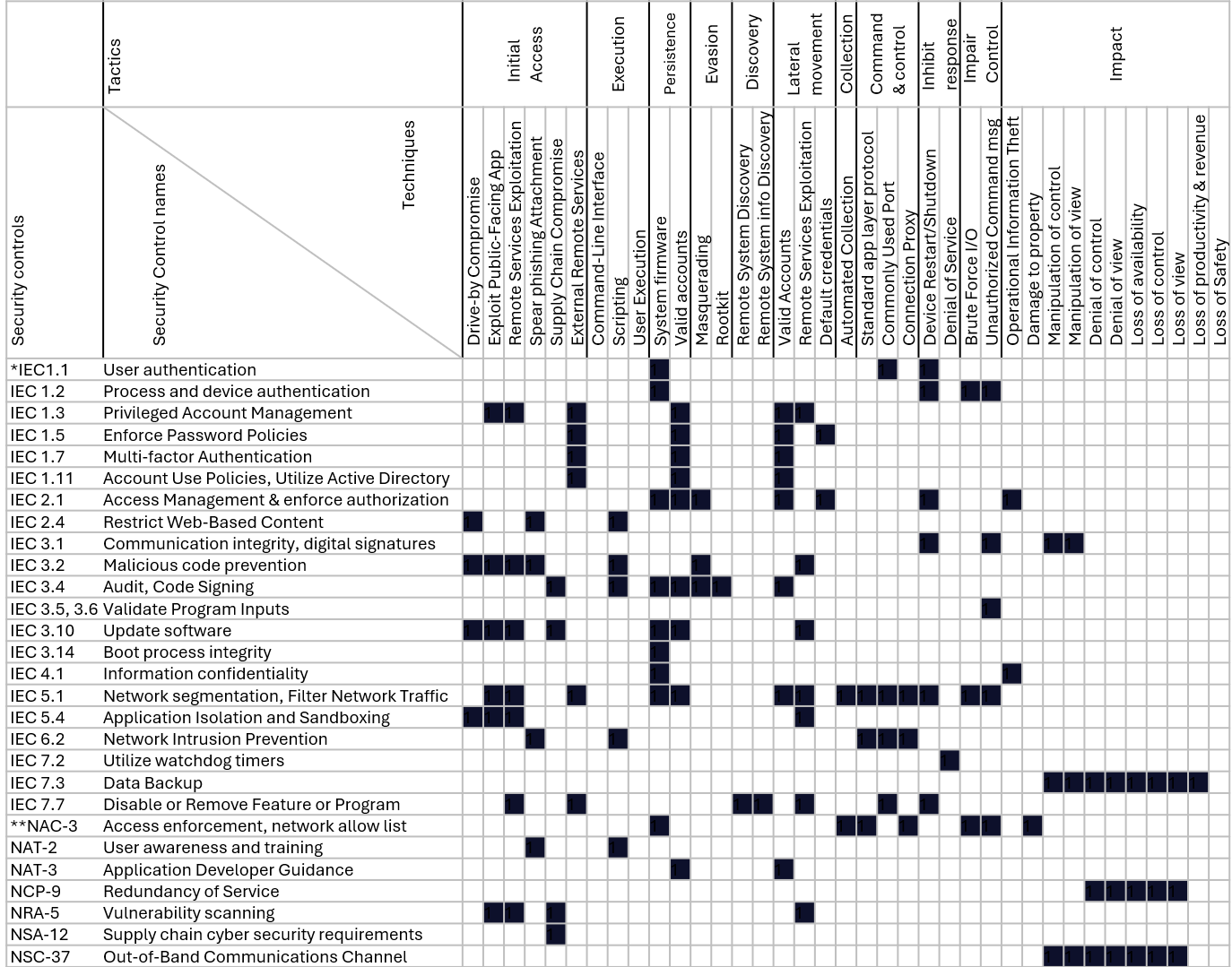}
	\caption{Security Controls, \\ Please note that to fit all the information inside the figure we had to use few abbreviations as follwos:\\ **IEC stands for IEC 62443-3-3:2013 - SR …., IEC 62443-4-2:2019 - CR ... ,  For instance, IEC 1.1. stands for IEC 62443-3-3:2013-SR 1.1,IEC 62443-4-2:2019-CR 1.1 \\ ***N stands for NIST SP 800-53 Rev. 4, For example, NAC-3 stands for NIST SP 800-53 Rev. 4 - AC-3. }
	\label{fig:security controls}
    
\end{figure}

To successfully detect and prevent an ICS incident, it is not sufficient to just understand the threats, it is also essential to defend against them. Hence, it is necessary to map ISA/IEC 62443 security controls to the MITRE ATT\&CK ICS framework. In some cases, where ISA/IEC 63443 controls don’t map, NIST 800-53 controls~\citep{29-nist_security_privacy_controls} are provided. In this work, the most common TTPs observed and reported by ICS threat intelligence organizations like Dragos are studied to efficiently prioritize the security controls required to counter the attacks. They are further grouped into three categories: (i) TTPs of attackers trying to gain initial access into OT and maintain their foothold in the OT environment, (ii) TTPs of attackers who have gained access into the OT environment and have attempted to gain higher level privileges to move through the ICS environment, and (iii) TTPs of attackers who have successfully caused an impact. Security controls to mitigate the risk are also mapped to these common techniques. The results are summarized in Figure~\ref{fig:security controls}, in which black cells indicate the specific security controls that have been utilized to address specific attack techniques.

\section{Discussion on future research}\label{sec6: future}
Implementing security controls requires proper testing and cannot be done in production environment, since it could result in shutdown of the process leading to business interruption. ICS testbeds has proven to be an effective approach for validating threat-based security controls in OT environment.

An industrial testbed is a controlled environment where a simulation or scaled-down version of an actual ICS is created to replicate real-world systems. Testbeds are primarily used to assess the security vulnerabilities of systems through replicating attacks. It can also be used to mitigate vulnerabilities by providing security controls to prevent the various attacks described above. There are two general methods for securing systems: i) secure-by-design or ii) secured after construction. 

Developing testbeds presents several challenges. The first is having a clear understanding of the architecture, devices, and protocols used in ICS implementations, which needs to accurately represent the real world and be replicable. Testbeds are complex and costly to configure and maintain, and a lack of documentation due to privacy concerns further complicates matters. 

Depending on the objective of the testbed and availability of resources, testbeds can be physical, virtual or hybrid. Physical testbeds involve realistic measurements, and the possibility of exploiting vulnerabilities, while virtual testbeds are low-cost simulations.

Even though a tremendous amount of effort has been made to simulate real-world incidents in a lab environment~\citep{30-conti_ics_testbeds}, testbeds utilizing the MITRE ATT\&CK framework to monitor and defend against common threats are not prevalent. The ATT\&CK framework utilizes a real-world dataset, which are essential to understand the evolving threat environment and implement security controls to mitigate it.

In this paper, we have taken the initial step to manually analyze the techniques used in well-known cyberattacks and identify the corresponding controls to address them. A possible direction for future research is to leverage larger datasets and employ automated methods, such as machine learning, to investigate the most frequently used techniques and identify effective mitigations in extensive historical data.

\section{Conclusion }\label{sec7: conlusion}
This work has examined the most common TTPs used by the threat actors by applying the MITRE ATT\&CK framework and has identified important security controls to defend against the reported threats. A brief overview of the ICS security reference architecture used widely in the industry, along with the evolution of the safety system architecture and related security issues have been discussed to provide a logical reasoning around the need for a threat-based approach to improve ICS security. 

Our analysis has identified that most cyber-attacks on OT systems originated in IT systems and the most common impact has been the shutdown or disruption of OT services due to targeted IT attacks. Based on the real-world incidents, we have then classified the TTPs of attackers into three categories: (i) attempts to gain initial access into OT and then maintain a foothold in the OT environment, (ii) gained access into the OT environment followed by the attempt to gain higher level privileges to move through the ICS environment and (iii) attacks with successful impact.

Many security controls that were identified are fundamental to establish ICS security practices. For example, network segmentation is an effective security control to prevent 12 different techniques used by attackers across the entire framework. Twenty-seven threat-based security controls have been identified in this work. To further refine and validate the critical controls needed to improve ICS security, ICS testbeds utilizing the MITRE ATT\&CK framework are required.

\bibliographystyle{unsrtnat}
\bibliography{references}  






\end{document}